\documentclass[aps. a4paper, twocolumn, nofootinbib, showpacs, amsfonts, amsmath,amssymb
]{revtex4}

\usepackage{graphicx}
\usepackage{dcolumn}
\usepackage{bm}

\usepackage{hyperref}

\def\be{\begin{equation}}
\def\ee{\end{equation}}
\def\bea{\begin{eqnarray}}
\def\eea{\end{eqnarray}}
\def\nn{\nonumber}

\begin{document}

\title{Intermediate mass ratio inspirals in dark matter halos}

\author{Darkhan Shadykul}
\email{darkhan.shadykul@nu.edu.kz}
\affiliation{Department of Physics, School of Sciences and Humanities, Nazarbayev University, Kabanbay Batyr 53, 010000 Astana, Kazakhstan}

\author{Hrishikesh Chakrabarty}
\email{hrishikesh.chakrabarty@nu.edu.kz}
\affiliation{Department of Physics, School of Sciences and Humanities, Nazarbayev University, Kabanbay Batyr 53, 010000 Astana, Kazakhstan}

\author{Daniele Malafarina}
\email{daniele.malafarina@nu.edu.kz}
\affiliation{Department of Physics, School of Sciences and Humanities, Nazarbayev University, Kabanbay Batyr 53, 010000 Astana, Kazakhstan}

\date{\today}

\begin{abstract}
 
We study eccentric equatorial orbits of a stellar-mass black hole around an intermediate-mass slowly rotating Kerr black hole in the presence of gravitational radiation and a dark matter halo. The stellar-mass companion will inspiral towards the central black hole while emitting gravitational waves.  
The evolution is affected by the dynamical friction force caused by the presence of dark matter surrounding the intermediate-mass black hole at the core. Previous studies have shown that the presence of dark matter can be deduced from the gravitational waveforms. 
We explore the combined effects of slow-rotation of the central black hole and the presence of the dark matter halo around it on the system's gravitational wave emission.

\end{abstract}

\maketitle

\section{\label{sec:level1}Introduction}

In 2015 the LIGO collaboration detected the first gravitational wave signal from the inspiral of two stellar mass black holes (BH) \cite{Abbott_2016}. Since then gravitational wave (GW) experiments have detected dozens of such events from binary mergers of black holes up to 142 solar masses. 
However current experiments are not able to detect gravitational wave signals from inspirals of stellar mass black holes into larger, more massive objects.
It is expected that future space-based detectors, such as LISA \cite{LISA}, TianQin \cite{Luo_2016}, or Taiji \cite{Taiji}, will be able to detect lower frequency events due to the emission of gravitational radiation from the capture of compact objects of stellar mass by massive black holes \cite{Colpi:2024xhw}.

It is also expected that the environment surrounding the central BH will play a crucial role in the exact form of the gravitational waves from inspirals onto supermassive and intermediate mass black holes (IMBHs). In particular, dark matter (DM) spikes should form around such black holes, thus affecting the orbital parameters of the inspiral orbits. 
In fact, it may even be possible to obtain information regarding the properties of DM from the study of the corresponding gravitational wave profiles \cite{Eda:2013gg,Macedo:2013qea, Barausse_2014, Hannuksela:2019vip,Figueiredo:2023gas,Rahman:2023sof,Karydas:2024fcn,Khalvati:2024tzz}.

The inspiral motion of a test particle around a massive Schwarzschild or Kerr black hole without DM halo has been extensively studied \cite{Amaro_Seoane_2006, mandel2008,hughes_2000}. It has been shown that the eccentricity of the secondary object's orbit decreases in both scenarios. However, under the scenario of a Kerr black hole, the eccentricity experiences a resurgence as the radius approaches the innermost stable circular orbit (ISCO).
More recently black hole mimickers, such as objects with non-Kerr multipole moments have also been considered \cite{Ryan:1995wh, Moore:2017lxy, Destounis:2023gpw}.

Several recent studies have focused on binary inspirals in the presence of a DM halo surrounding the primary object, given by a non-rotating Schwarzschild black hole \cite{Eda_2015,Yue_2019,Kavanagh_2020,Becker_2022,Coogan_2022, Dai_2022, Destounis:2022obl,Zhao:2024bpp,Li:2021pxf,Kadota:2023wlm,Kim:2024rgf}. According to the studies, the orbital eccentricities of a secondary object tend to increase in the presence of DM due to the inverse proportionality of the dynamical force to the square of the velocity \cite{Yue_2019}. On the other hand, the feedback on the DM spike might also affect the size of the dynamical friction force \cite{Mukherjee:2023lzn,Karydas:2024fcn,Kavanagh:2024lgq}.
Note that the dynamical friction need not be the result of DM alone, as it is produced also by distributions of ordinary matter. Therefore, knowledge of the BH's environment in terms of normal matter is necessary in order to distinguish the DM effects.
Inspirals in a perturbed Schwarzschild and Kerr metric to account for the presence of external matter fields were considered in \cite{Polcar:2022bwv,Ghosh:2024arw}. There have also been studies on gravitational waves from hyperbolic encounters in the presence of DM \cite{AbhishekChowdhuri:2023rfv}. 

In the present article we consider eccentric equatorial orbits of a stellar mass black hole around an intermediate mass Kerr black hole immersed in a DM halo, in the presence of gravitational radiation and dynamical friction force. In particular, we focus on intermediate mass ratio inspiral (IMRI) systems, with a mass ratio of $\sim10^4$. We consider a massive central spinning black hole with a solar mass orbiting companion traveling in its gravitational field. Once in a bound orbit, the smaller compact object will inspiral towards the core black hole while emitting gravitational radiation. 

The intermediate mass ratio nature of the system means that the generated gravitational waveforms can be determined accurately using black hole perturbation theory \cite{Teukolsky:1973ha}. However, such calculations are computationally very expensive. In this article we use the ``kludge'' method \cite{Babak:2006uv} which is a way of generating approximate waveforms that still captures the main features of  the fully relativistic waveforms \cite{Babak:2006uv,Glampedakis_2002,Gair:2005ih}. In this method, the orbital dynamics are fully relativistic at each instant however the energy and angular momentum fluxes evolving the system from one geodesics to the next are approximate and Newtonian. The primary use of kludge waveforms is to generate mock data which then can be used to develop further tools for statistical analyses \cite{Arnaud:2006gm,MockLISADataChallengeTaskForce:2009wir}. Note that, while modeling these systems in the relativistic regime as Newtonian will lead to significant biases in parameter estimation, it is still useful in order to study the qualitative behavior of the system \cite{Babak:2006uv}.

We consider adiabatic motion where the timescale of the inspiral is much greater than the time period of the orbital motion. We include the presence of DM surrounding the core IMBH which causes dynamical friction force, which in turn affects the dynamics of the inspiral.
Our objectives are twofold: (a) investigate the combined effects of the black hole spin and the presence of DM halo on the gravitational wave emission and (b) gain insights on how the eccentricity and the other orbital parameters of the inspiralling object are affected by the central black hole spin, as compared to previous studies. 
Given the inverse relationship between the frequency of emitted waves and the central black hole's mass, it is likely that they will fall within the low-frequency range of $10^{-5}$ to $10^{-1}$ Hz, which is where LISA's sensitivity is highest. Therefore, it is possible that the detection of GWs from such systems could provide information on the nature of DM. 

The paper is organized as follows: In section \ref{sec:level2} we outline the main features of the model, namely the geometry of the central black hole, the DM models considered, the orbits of the smaller compact object, dynamical friction and gravitational wave emission. In section \ref{results}, we describe the model's setup with the chosen initial values of the parameters and outline the results with particular emphasis on the effects of the presence of DM. Finally in Sec.~\ref{conc}, we summarize our findings. Throughout this paper, we have adopted geometrized units $c = G = 1$.

\section{\label{sec:level2}Dynamic IMRI model}
Our aim is to study eccentric equatorial orbits of a stellar mass black hole around an intermediate-mass Kerr black hole in the presence of gravitational radiation and dynamic friction force. 
In the following we shall briefly review all the necessary ingredients of our model.

\subsection{Kerr black hole}
The Kerr metric in Boyer-Lindquist coordinates is given by \cite{Kerr:1963ud,Boyer_1967}
\begin{eqnarray}
\label{eq:Kerr}
 ds^2 = && -\left( 1- \frac{r_sr}{\Sigma} \right) dt^2 + \frac{\Sigma}{\Delta}dr^2 -\frac{2r_sra}{\Sigma}\sin^2 \theta dt d\phi\nonumber +\\
 && +\Sigma d\theta^2 + \left(r^2 + a^2 + \frac{r_sra^2 \sin^2 \theta}{\Sigma} \right) \sin^2 \theta d\phi^2 ,
\end{eqnarray}
where $r_s = 2m_1$ is the Schwarzschild radius with $m_1$ being the black hole mass, $a = J/m_1$ is the Kerr spin parameter, $\Sigma = r^2 + a^2\cos^2 \theta$, and $\Delta = r^2 + a^2 - 2r_sr$. As it is well known, the Kerr geometry is fully characterized by two quantities, namely the black hole's mass $m_1$ and the spin parameter $a$, which is one of our main input variables, as it not only characterizes the rate of rotation of the black hole but also affects the trajectories of smaller orbiting bodies.

The Kerr spacetime is completely integrable and therefore timelike geodesics for massive test particles (i.e. test particles with mass $m_2 \ll m_1$) posses four independent conserved quantities, namely the energy per unit mass $\mathcal{E}$, the angular momentum per unit mass projected along the black hole’s spin axis $L_z$, the Carter constant per unit mass squared $Q$ \cite{Carter_1968}, and the 4-momentum normalization:
\bea
\label{eq:E}
 \mathcal{E} &=& - u_0 = - g_{00}u^0 - g_{0\phi}u^{\phi}, \\
\label{eq:L}
 L_z &=& - u_{\phi} = g_{0\phi}u^0 + g_{\phi\phi}u^{\phi}, \\
\label{eq:C}
 Q &=& \Sigma^2(u^{\theta})^2 + \sin^{-2}\theta L_z^2 + a^2 \cos^2 \theta(1-\mathcal{E}^2), \\
\label{eq:p}
 - m_2^2 &=& g_{\mu\nu}p^{\mu}p^{\nu} .
\eea

Geodesics of test particles in the Kerr spacetime can be parameterized using three orbital elements, namely the inclination angle $\iota$, the orbital eccentricity $e$, and the semi-latus rectum $p$ (or alternatively the semi-major axis $\bar{a}$). 
In the following we will consider only equatorial orbits, 
for which Carter constant may be expressed in terms of the angular momentum $L_z$, thus setting $\iota =0$.

One of the interesting properties of geodesic motion of test particles in the Kerr spacetime is that, due to its axial symmetry, geodesics that start in the equatorial plane would stay in the equatorial plane at later times as well. 
A very relevant quantity for us (and of astrophysical interest, in general) would be the location of the innermost stable circular orbit (ISCO). The conserved quantities in Eq.~\eqref{eq:E} and \eqref{eq:L} can be used to reduce the problem of motion in the equatorial plane to a one dimensional first order equation of the form $\dot{r}^2+V_{\rm eff}(r)={\rm const}.$, where $V_{\rm eff}$ is the effective potential for the geodesic motion for test particles. 
Then, the ISCO can be found from the equation $V_{\rm eff}''(r) = 0 $, where the prime indicates derivative with respect to the radial coordinate. In case of the static, spherically symmetric spacetime, i.e. the Schwarzschild metric, the ISCO is at $r = 6m_1$. However, in the Kerr metric, the expression for the same is quite complicated and depends on the spin parameter $a$. The equation $V_{\rm eff}''(r) = 0 $ in the Kerr metric admits two solutions: one corresponds to corotating or prograde (i.e. rotation along the BH's spin direction) orbits and the other to counterrotating or retrograde (i.e. rotation against the BH's spin direction) orbits. For $a = 0$, the two solutions coincide at $r=6m_1$. With increasing $a$, the ISCO moves closer to the BH for corotating particles and farther for counterrotating particles. At the extremal value of the spin, i.e. at $a = 1$, the corotating ISCO coincides with the outer horizon of the Kerr BH \cite{Bardeen:1972fi}.

\subsection{Dark matter spike}

An intermediate mass black hole may form a DM mini-halo under certain black hole formation scenarios \cite{Navarro_1996,Zhao_2005,Bertone_2005}. According to current models the halo is redistributed into a cusp by the black hole, which is referred to as the `spike'. This was first suggested by Gondolo and Silk \cite{Gondolo_1999}, who demonstrated using Newtonian analysis that a spike in the density of DM forms in the vicinity of a black hole. 
In \cite{Sadeghian_2013}, a comprehensive relativistic computation was conducted for a spherical black hole. The results showed that the spike attains considerably greater densities and extends closer to the event horizon.

DM spikes surrounding a Kerr black hole were investigated in \cite{Ferrer_2017}. The study compared the characteristics of the `cuspy' Hernquist profile surrounding Schwarzschild and Kerr black holes. It was shown that the density profile of DM is affected by the spin parameter and it is steeper for Kerr black holes with higher angular momentum. However, this difference becomes significant only for spin parameter $a \gtrsim 0.5$, and at distances $r \lesssim 2.5 r_{\text {isco}}$, as illustrated by the plot in Fig. 10 in \cite{Ferrer_2017}. We aim to consider DM profiles that neglect the effects of the black hole's angular momentum.
Therefore, in view of the results obtained in \cite{Ferrer_2017}, our focus in the following will be on slowly rotating black holes with a spin parameter $0<a<0.3$. 
This allows us to utilize the DM profiles of a Schwarzschild black hole without having to consider the effects of the black hole's spin on the DM particles. On the other hand, the relativistic effects on the DM profile is only relevant in the innermost region of the halo where the DM density extends all the way to $ r = 2r_s $ \cite{Sadeghian_2013}, instead of vanishing at $4r_s$ as found earlier in \cite{Gondolo_1999}. In addition, the total mass of DM particles enclosed in the innermost region is small and hence we expect that the additional perturbing effects on the orbit of the secondary to be small as well \cite{Sadeghian_2013}. We defer the examination of more general cases of Kerr black hole and DM profiles with relativistic effects to future investigations. Also we shall mostly restrict the analysis to distances from the central black hole large enough to justify the above choice of DM profile. 

The adiabatic growth of a black hole at the center of a DM halo with a singular power-law density $\rho(r) \propto r^{-\alpha_{ini}}$ leads to the emergence of a high density DM region with $\rho(r) \propto r^{-\alpha}$ with $2.25 \leq \alpha \leq 2.5$.
The DM minispike around an IMBH can then be modeled with a spherically symmetric distribution that follows a single power law

\begin{equation}
\label{eq:DM spike}
\rho_{\text{dm}} = \begin{cases} \rho_{\text{sp}} \left( \frac{r_{\text{sp}}}{r} \right)^{\alpha} & r_{\text{in}} \leq r \leq r_{\text{sp}}  \\ 0. & r<r_{\text{in}} \end{cases}
\end{equation}

where $r_{\text{sp}}$ is the characteristic radius of the DM minispike and $\rho_{\text{sp}}$ is the DM density at the radius $r_{\text{sp}}$. The inner radius is taken as $r_{\text{in}} = 4m_1$ \cite{Sadeghian_2013}. According to different models the power law index range is $1 < \alpha < 3$. 
In the following we consider three widely used models: 
self-interacting dark matter (SIDM) with $\alpha = 7/4$ \cite{Shapiro_2014}, Navarro, Frenk and White (NFW) with $\alpha = 7/3$ \cite{Navarro_1996} and DM spike around primordial black holes with $\alpha = 9/4$ \cite{Boudaud_2021}. 

\begin{figure}[h]
\centering
\includegraphics[scale=0.25]{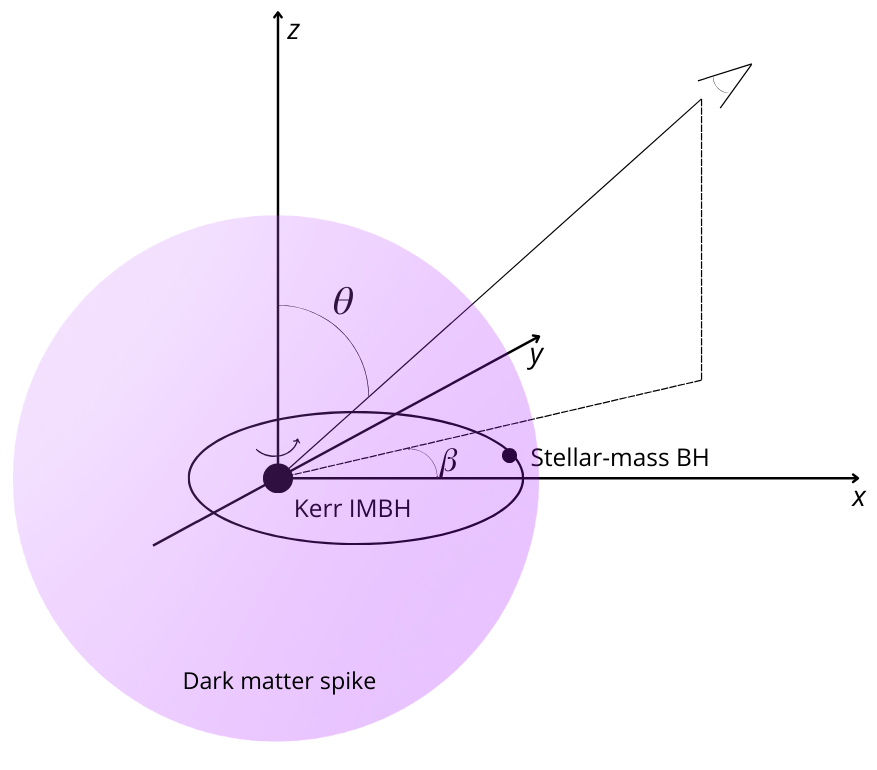}
\caption{A stellar-mass black hole with mass $m_2$ orbits an intermediate-mass Kerr black hole with mass $m_1$ on a Keplerian orbit in the equatorial plane, in the presence of a dark matter halo (light pink region). The inclination angle of the line of sight of a distant observer with respect to the normal to the equatorial plane is denoted by $\theta$, while $\beta$ is the angle in the orbital plane between the major axis and the observer's direction.}
\label{fig:dmhalo}
\end{figure}

SIDM has been considered as an alternative to collisionless cold dark matter to model DM distributions around black holes \cite{Shapiro_2014}.
The NFW profile is a universal density profile for DM halos that was derived
via cosmological N-body numerical simulations \cite{Navarro_1996}. 
On the other hand, primordial black holes may have been produced in the early universe after cosmic inflation. In that case, DM in the form of particles can be subsequently accreted around these objects, in particular when it gets non-relativistic. This may result in a DM mini-spike around each black hole, with density orders of magnitude larger than the cosmological one \cite{Boudaud_2021}.

For the purposes of our work, a  description of a static halo as given in \ref{eq:DM spike} would be insufficient. It is essential to account for the distribution of DM particles not only in terms of their spatial distance but also in relation to their varying velocities.
Given a spherically symmetric and isotropic DM halo, the DM particles in the spike can be described by an equilibrium phase space distribution function 
\cite{Kavanagh_2020}

\begin{equation}
\label{eq:distribution}
 f(\mathcal{E}) = m_{\text{DM}}\frac{dN}{d^3\textbf{r}d^3\textbf{v}},
\end{equation}
where $m_{\text{DM}}$ is the mass of a DM particle and
\begin{equation}
 \mathcal{E} = \Psi(r) - \frac{1}{2}v^2,
\end{equation}
is the relative energy per unit mass, while $\Psi(r)$ is the gravitational potential. Gravitationally bound particles correspond to those with $\mathcal{E}>0$. 
Neglecting the DM halo's contribution to the gravitational potential for small distances from IMBH, the potential can be taken as $\Psi(r) = m_1/r$. 

With a given density profile $\rho(r)$ the distribution function can be recovered using Eddington's inversion formula \cite{binney2011galactic}.

\begin{equation}    
\label{eq:Eddington}
 f(\mathcal{E})= \frac{1}{\sqrt{8}\pi^2} \int_0^\mathcal{E} \frac{d\Psi}{\sqrt{\mathcal{E}-\Psi}} \frac{d^2\rho}{d\Psi^2}
\end{equation}

For a power-law spike it can be expressed as (see Appendix in \cite{Edwards_2020})

\begin{equation}    
\label{eq:distr}
 f(\mathcal{E})= \frac{\alpha(\alpha - 1)}{(2\pi)^{3/2}}\rho_{\text{sp}}\left(\frac{r_{\text{sp}}}{m_1} \right)^{\alpha} \frac{\Gamma(\alpha - 1)}{\Gamma(\alpha - \frac{1}{2})} \mathcal{E}^{\alpha - 3/2},
\end{equation}
where $\Gamma (x)$ is the gamma function. For a given distribution function the corresponding density function can be found by solving the following integral
\begin{equation}
    \label{eq:rhof}
    \rho(r) = 4\pi \int_0^{v_{\text{max}}(r)}v^2 f \left( \Psi(r) - \frac{1}{2}v^2 \right) dv ,
\end{equation}
where $v_{\text{max}}(r) = \sqrt{2\Psi(r)}$ is the escape velocity.

\subsection{Keplerian orbit}
We utilize an approximate method (the so called ``Kludge" method), developed by Glampedakis et al. \cite{Glampedakis_2002} for the construction of inspirals of stellar-mass companions orbiting an intermediate mass Kerr black hole in the presence of gravitational radiation. This method constructs the trajectory of the companion from an exact Kerr geodesic, while the orbital parameters are evolved using expressions derived from post-Newtonian expansions.
Although the kludge method is inferior to the fully relativistic numerical and analytical methods, the fluxes and the gravitational waveforms calculated this way still capture important qualitative features at a fraction of the computational cost \cite{Babak:2006uv}. We stress that, for a more self-consistent model, one needs to consider fully relativistic waveforms incorporating black hole perturbation theory.

Here we concentrate on prograde motion around a Kerr black hole, when the inclination angle of the secondary object's orbit is taken to be $\iota = 0^{\circ}$. The inclination angle is defined as \cite{Glampedakis_2002}:
\begin{equation}
\label{inclination}
 \cos \iota = \frac{L_z}{\sqrt{Q+L_z^2}},
\end{equation}
so that $L_z=L$.
The model 
assumes an adiabatic approximation to calculate the rates of change of semi-latus rectum $\dot{p}$ and eccentricity $\dot{e}$ using the weak-field quadrupole-order fluxes for energy $\dot{E} $ and angular momentum $ \dot{L}$.
Here, the dot denotes derivatives with respect to the coordinate time $t$.
The gravitational waveform for distant observers is then evaluated by relating the Boyer-Lindquist coordinates of the Kerr spacetime to spherical coordinates in flat space.

The IMRI system consists of two black holes, one of which has a much larger mass than the second, $m_1 \gg m_2$. Particularly, we consider a system with an IMBH with $m_1 \simeq 10^5 M_{\odot}$ in the center and a secondary object, a stellar mass black hole, with $m_2 \simeq 10 M_{\odot}$. Due to the large mass difference, the secondary black hole rotates around the first in a Keplerian orbit. The reduced mass is nearly equal to $m_2$ and the barycenter of the system is approximately equivalent to the position of the IMBH.

We assume that the total DM mass around the IMBH is much less than the central black hole's mass. Therefore we may ignore the DM mass when calculating the total mass of the system and the reduced mass. This is especially likely to be relevant if we consider low eccentricity orbits.

The secondary object loses energy both due to the radiation of gravitational waves and dynamical friction by DM particles, which leads to an inspiral orbit \cite{Chandrasekhar_1943} with
\bea
\label{eq:energy}
\frac{dE}{dt}  &=& \left\langle \frac {dE_{GW}}{dt} \right\rangle + \left\langle \frac {dE_{DF}}{dt} \right\rangle , \\
\label{eq:angular momentum}
\frac{dL}{dt}  &=& \left\langle \frac {dL_{GW}}{dt} \right\rangle + \left\langle \frac {dL_{DF}}{dt} \right\rangle.
\eea

The orbital energy and the angular momentum relations are given by \cite{Maggiore_2007} 
\bea
\label{eq:total energy}
E  &=&  -\frac {m\mu}{2\bar{a}} , \\
\label{eq:ecc l}
e^2 -1  &=&  \frac {2EL^2}{m^2\mu^3} .
\eea
where $\bar{a}$ and $e$ are the semi-major axis and eccentricity of the orbits respectively, while $m = m_1 + m_2$ and $\mu = m_1m_2/(m_1+m_2)$ are the total mass and the reduced mass of the system respectively.
As a consequence of energy and angular momentum loss caused by gravitational radiation, integrals of motion change. They should undergo adiabatic change over timescales significantly longer than any orbital timescale when the mass ratio is small.

In the following we limit to the weak-field approximation which adequately characterizes orbits for which the semi-latus rectum $p=\bar{a}(1-e^2)$ is very large, $p/m_1\gg 1$. Specifically, the energy and angular momentum fluxes should be expressed in quadrupole-order formulae to provide adequate precision.
The average time derivatives of the energy and the angular momentum loss due to the gravitational wave emission for a Kerr black hole in the leading post-Newtonian order are expressed as \cite{Glampedakis_2002}

\begin{eqnarray}
\label{eq:energy gw}
 \dot{E}_{GW} &= &-\frac {32}{5} \frac{m_2^2}{m_1^2} \left( \frac{m_1}{p} \right)^5 (1-e^2)^{3/2} \nonumber\\
&&\times \left[f_1(e) - \frac {a}{m_1} \left( \frac{m_1}{p} \right)^{3/2} f_2(e) \right], \\
\label{eq:angular gw}
 \dot{L}_{GW} &= &-\frac {32}{5} \frac{m_2^2}{m_1^2} \left( \frac{m_1}{p} \right)^{7/2} (1-e^2)^{3/2} \nonumber\\
 &&\times \left[f_3(e) + \frac {a}{m_1} \left( \frac{m_1}{p} \right)^{3/2} (f_4(e) - f_5(e)) \right],
\end{eqnarray}
where $f_1, ... , f_5$ are given by the following expressions
\bea
\label{eq:f1}
    f_1(e) &=& 1 + \frac{73}{24}e^2 + \frac{37}{96}e^4 , \\
\label{eq:f2}
    f_2(e) &=& \frac{73}{12} + \frac{823}{24}e^2 + \frac{949}{32}e^4 + \frac{491}{192}e^6 , \\
\label{eq:f3}
    f_3(e) &=& 1 + \frac{7}{8}e^2, \\
\label{eq:f4}
    f_4(e) &=& \frac{61}{24} + \frac{63}{8}e^2 + \frac{95}{64}e^4 , \\
\label{eq:f5}
    f_5(e) &=& \frac{61}{8} + \frac{91}{4}e^2 + \frac{461}{64}e^4 .
\eea
These fluxes were derived by Glampedakis et al. \cite{Glampedakis_2002}, based on previous studies by Ryan \cite{Ryan_1996}. 

\subsection{Dynamical friction}
In addition to energy and angular momentum loss due to gravitational waves, when a stellar mass object travels through a DM environment, it undergoes gravitational interactions with the DM particles in the halo, resulting in dynamical friction, leading to additional loss of energy and angular momentum. 
The Chandrasekhar dynamical friction \cite{Chandrasekhar_1943} is given by
\begin{equation}
\label{eq:force}
 F_{DF}(r) = \frac{4\pi m_2^2 \rho_{\text{dm}}(r) \xi(v) \log \Lambda}{v^2},
\end{equation}
where $\log \Lambda$ is the Coulomb logarithm and can be taken as \cite{Kavanagh:2020cfn} (see also \cite{Kavanagh:2024lgq} for some recent insights)
\begin{equation}
\label{eq:log lambda}
 \log \Lambda = \log \sqrt{m_1/m_2}.
\end{equation}

The term $\xi(v)$ denotes the fraction of DM particles moving slower than the orbital speed of the secondary object. In other words, we should replace the upper limit in Eq.~\eqref{eq:rhof} by the speed of the secondary $v_{\text{orb}}$.
\begin{equation}
    \label{eq:rhov}
    \rho(r) \xi(v) = 4\pi \int_0^{v_{\text{orb}}(r)}v^2 f \left( \Psi(r) - \frac{1}{2}v^2 \right) dv 
\end{equation}

The DM particles that move slower than the secondary object will cause a decrease in its kinetic energy (that is, ``cooling''), while those particles that move faster will cause ``heating''. The ratio between heating and cooling is proportional to the ratio of the mass of a DM particle to the orbiting BH mass $m_{\text{DM}}/m_2$, therefore, the heating can be neglected (see \cite{binney2011galactic} at p.582, \cite{Kavanagh_2020} and \cite{Chandrasekhar_1943}) and the net friction force depends only on the density of DM particles moving more slowly than $v_{\text{orb}}$ of the secondary BH.

The average energy and momentum losses due to dynamical friction are expressed as \cite{Yue_2019,Becker_2022}
\bea
\label{eq:Edf}
 \left \langle \frac{dE_{DF}}{dt} \right \rangle  &=& \frac{1}{T} \int \limits_0^T \frac{dE}{dt}dt = \int \limits_0^T F_{DF}dt, \\
\nn
 \left \langle \frac{dL_{DF}}{dt} \right \rangle  &=& \frac{1}{T} \int \limits_0^T \frac{dL}{dt}dt = \\ 
 \label{eq:Ldf}
 &=& - \frac{\sqrt{ma(1-e^2)}}{T} \int \limits_0^T \frac{F_{DF}}{v}dt.
\eea

By differentiating equation (\ref{eq:total energy}) we obtain
\bea
\label{eq:deda}
 \frac{dE}{d\bar{a}} &=& \frac {m_1m_2}{2\bar{a}^2}, \\
\label{eq:dadt}
 \frac{d\bar{a}}{dt} &=& \frac {dE/dt} {dE/d\bar{a}}.
\eea
Similarly, for the eccentricity we can write
\begin{equation}
\label{eq:dedt}
 \frac{de}{dt} = - \frac{1-e^2}{2e} \left( \frac{dE/dt}{E} + 2 \frac{dL/dt}{L}  \right).
\end{equation}
We thus obtain a system of equations (\ref{eq:energy}), (\ref{eq:angular momentum}), (\ref{eq:energy gw}), (\ref{eq:angular gw}), (\ref{eq:dadt}), (\ref{eq:dedt}) that can be solved numerically.

\subsection{Gravitational waves}

Previous studies have demonstrated that the presence of DM can be detected by analyzing gravitational waveforms. One of the objectives of our study is to examine the possibility of detecting a Kerr black hole with a DM halo by analyzing the GW waveform.

Gravitational waves are generated by the binary system as a result of changes in its quadrupole moment. The two independent GW polarization modes are represented by the following expression
\cite{Martel_1999}:

\begin{widetext}
\bea
\label{eq:h+} \nn
 h_+ &=& - \frac{m \mu}{pD_L} \left[ \left( 2 \cos(2\phi - 2\beta) + \frac{5}{2}e \cos(\phi - 2 \beta) + \frac{1}{2} e \cos(3\phi - 2\beta) + e^2 \cos(2\beta) \right)(1 + \cos^2 \theta) + \right. \\ 
 &&\Big. +(e \cos \phi + e^2) \sin^2 \theta \Big], \\
\label{eq:hx}
 h_{\times} &=& - \frac{m \mu}{pD_L} \left[ 4 \sin(2\phi - 2\beta) + 5e \sin(\phi - 2 \beta) + e \sin(3\phi - 2\beta) - 2e^2 \sin(2\beta) \right] \cos \theta,
\eea
\end{widetext}
where $D_L$ is the luminosity distance to the source, $\theta$ is the inclination angle of the orbital plane with respect to the plane of the sky (see Fig.~\ref{fig:dmhalo}), $\phi$ is the true anomaly, and $\beta$ is the angle created in the orbital plane between the major axis and the observer's direction \cite{Moreno-Garrido_1995}.

The GW signal may be broken down into the harmonics of the mean orbital frequencies as follows \cite{Martel_1999}:
\begin{equation}
\label{hfourier}
 h_{+, \times} = \mathcal{A} \sum_{n=1}^{\infty} \left( C_{+,\times}^{(n)} \cos(nl) + iS_{+,\times}^{(n)} \sin(nl) \right).
\end{equation}

The coefficients $C_{+,\times}$ and $S_{+,\times}$ can be obtained using the Fourier-Bessel expansion of the orbital motion \cite{Moreno-Garrido_1995,Moore_2018,Chandramouli_2022}. This yields the coefficients
\begin{widetext}
\bea
\label{C+}
 C_{+}^{(n)} &=& \left[ 2\sin^2\theta J_n(ne) + \frac{2}{e^2}(1+\cos^2\theta)\cos{2\beta} \left( (e^2 - 2) J_n(ne) + ne(1-e^2)(J_{n-1}(ne) - J_{n+1}(ne)) \right) \right], \\
\label{S+}
 S_{+}^{(n)} &=& -\frac{2}{e^2} \sqrt{1-e^2}(1+\cos^2{\theta}) \sin{2\beta} \left[ -2(1-e^2)nJ_n(ne) +e(J_{n-1}(ne) - J_{n+1}(ne)) \right], \\
\label{Cx}
 C_{\times}^{(n)} &=& - \frac{4}{e^2}\cos {\theta} \sin{2\beta} \left[ (2-e^2)J_n(ne)+ne(1-e^2)(J_{n-1}(ne)-J_{n+1}(ne)) \right], \\
\label{Sx}
 S_{\times}^{(n)} &=& -\frac{4}{e^2} \sqrt{1-e^2}\cos{\theta} \cos{2\beta} \left[ -2(1-e^2)nJ_n(ne) +e(J_{n-1}(ne) - J_{n+1}(ne)) \right],
\eea
where $J_n$ are Bessel functions of the first kind.
\end{widetext}

\section{Numerical results}\label{results}

\begin{figure}[tt]
\centering
\includegraphics[width=0.92\linewidth]{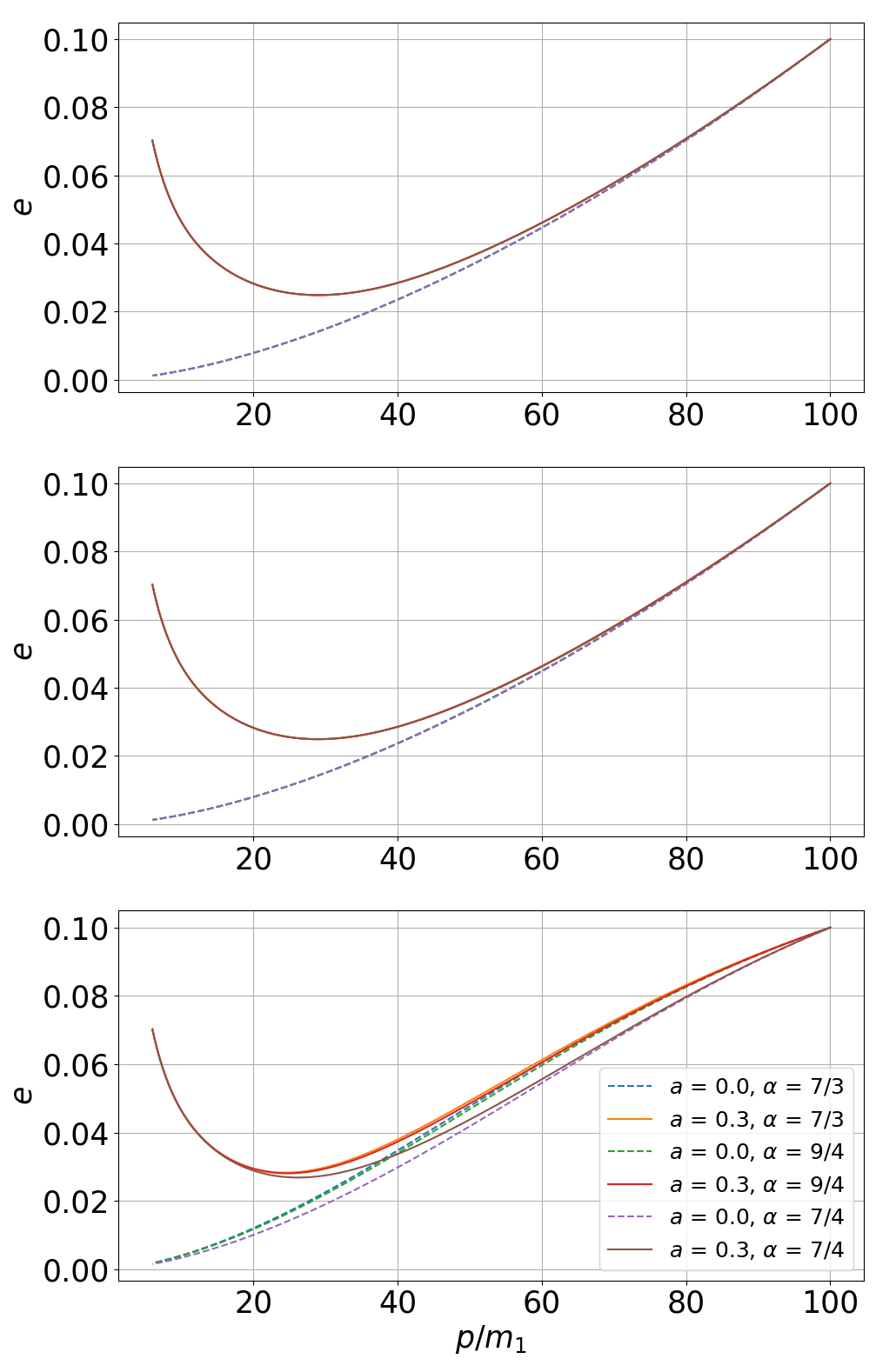}
\caption{The evolution of orbital eccentricity $e$ for a black hole spiraling towards a non-rotating (dashed) and a Kerr (solid) IMBH as a function of semi-latus rectum $p$. The initial eccentricity is $e_0=0.1$ and the initial semi-latus rectum is $p_0 = 100m_1$. The DM spike densities in the panels from top to bottom are $\rho_{\rm sp} = [0, \ 5.448\times10^{15}, \ 5.448\times10^{17}]$. The three models considered for the DM profile almost completely overlap for the lower density case (middle panel) and become distinguishable only for large $\rho_{\rm sp}$ (bottom panel).} 
\label{fig:fig1}
\end{figure}

\begin{figure}[h]
\centering
\includegraphics[width=1\linewidth]{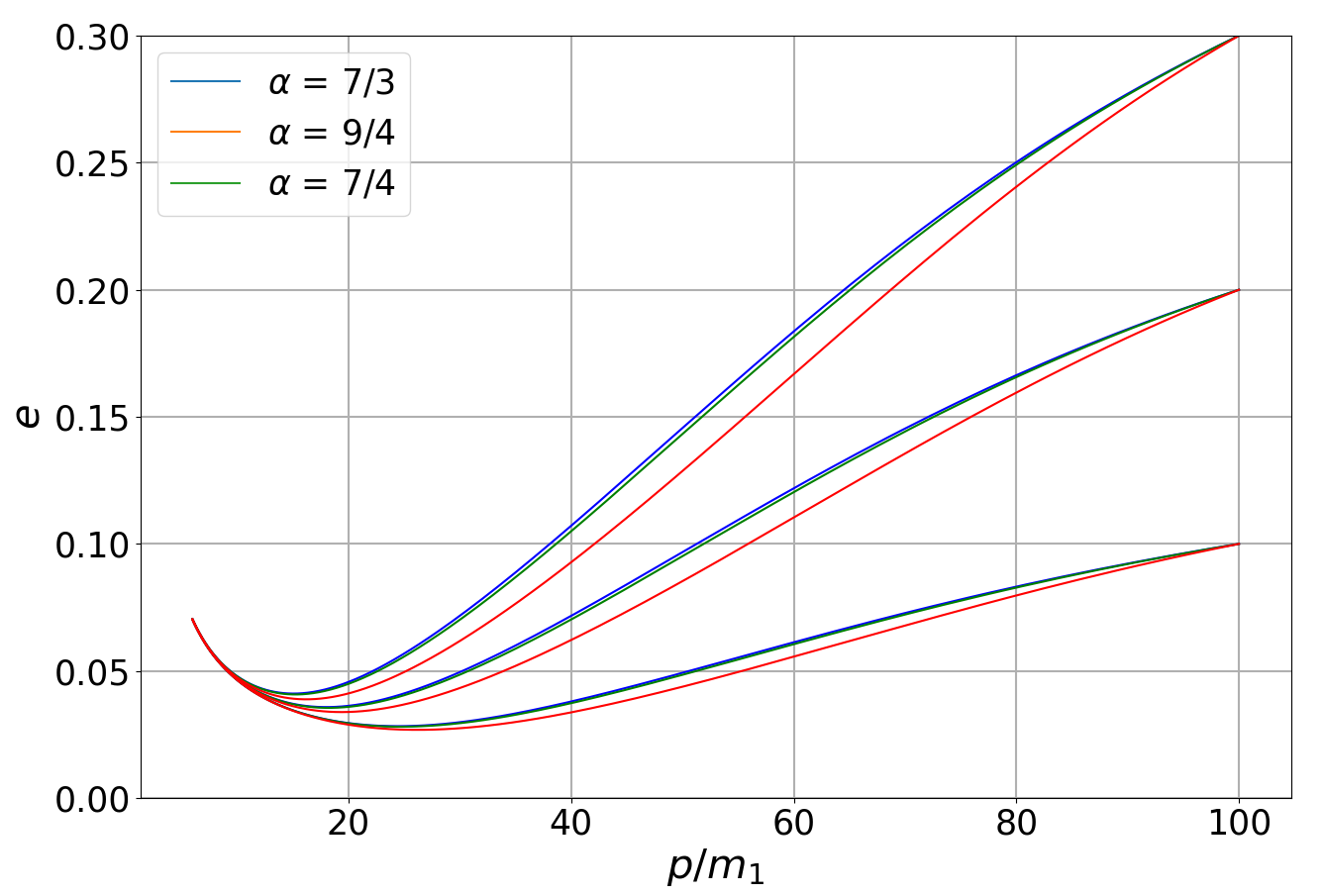}
\caption{The evolution of the eccentricity $e$ of inspirals onto a Kerr black hole with $a=0.3$ for different initial values of $e_0 = [0.1, \ 0.2, \ 0.3]$. 
The DM spike density here is $\rho_{\rm sp} = 5.448\times10^{17}$. 
}
\label{fig:fig5}
\end{figure}

\begin{figure}[tt]
\centering
\includegraphics[width=0.91\linewidth]{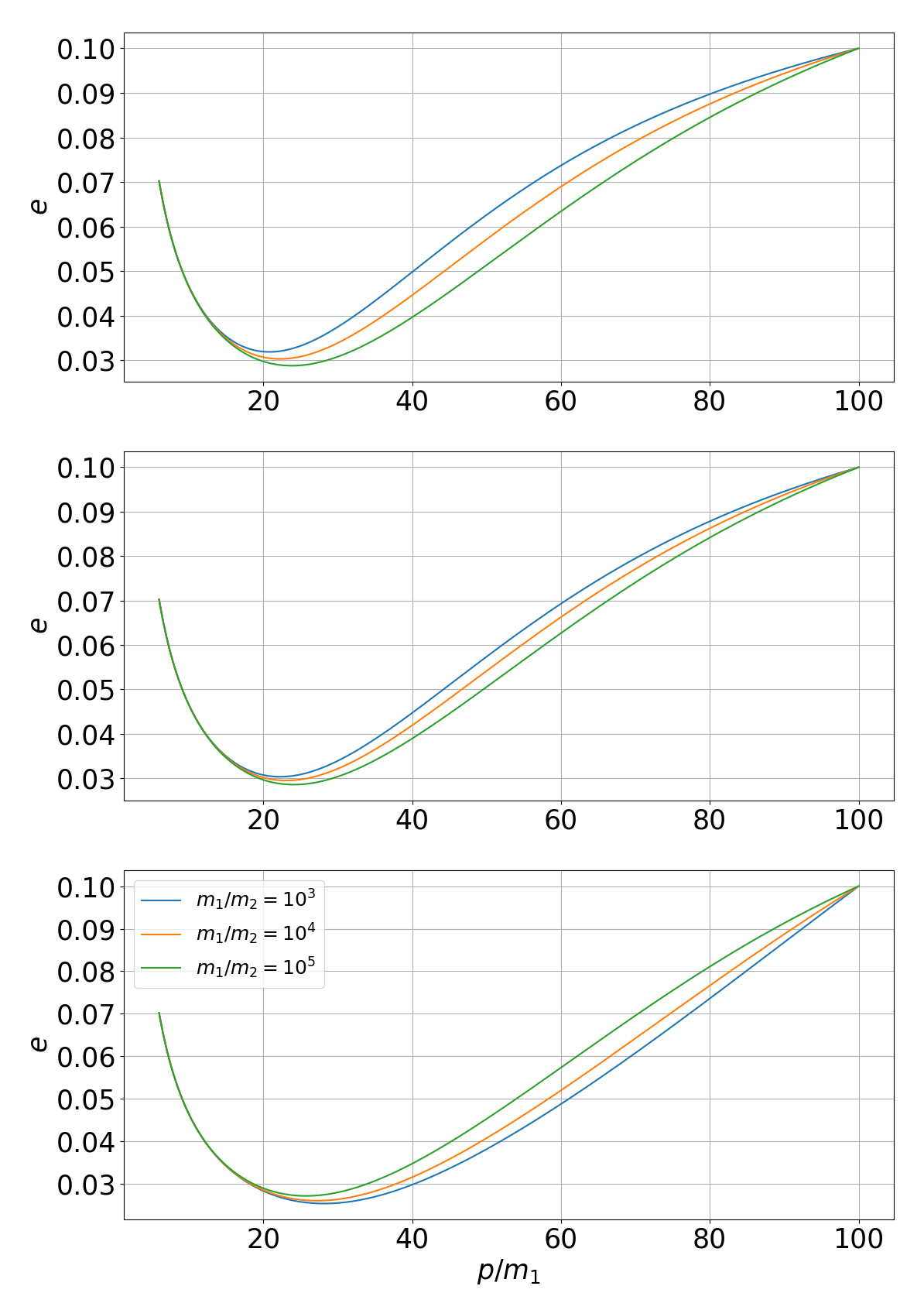}
\caption{The evolution of orbital eccentricity $e$ for a black hole spiraling towards a Kerr IMBH as a function of semi-latus rectum $p$ for different mass ratios. The initial eccentricity is $e_0=0.1$, and the initial semi-latus rectum is $p_0 = 100m_1$. The DM spike powers in the panels from top to bottom are $\alpha = [7/3, 9/4, 7/4]$.} 
\label{fig:fig6}
\end{figure}

\begin{figure}[tt]
\centering
\includegraphics[width=0.91\linewidth]{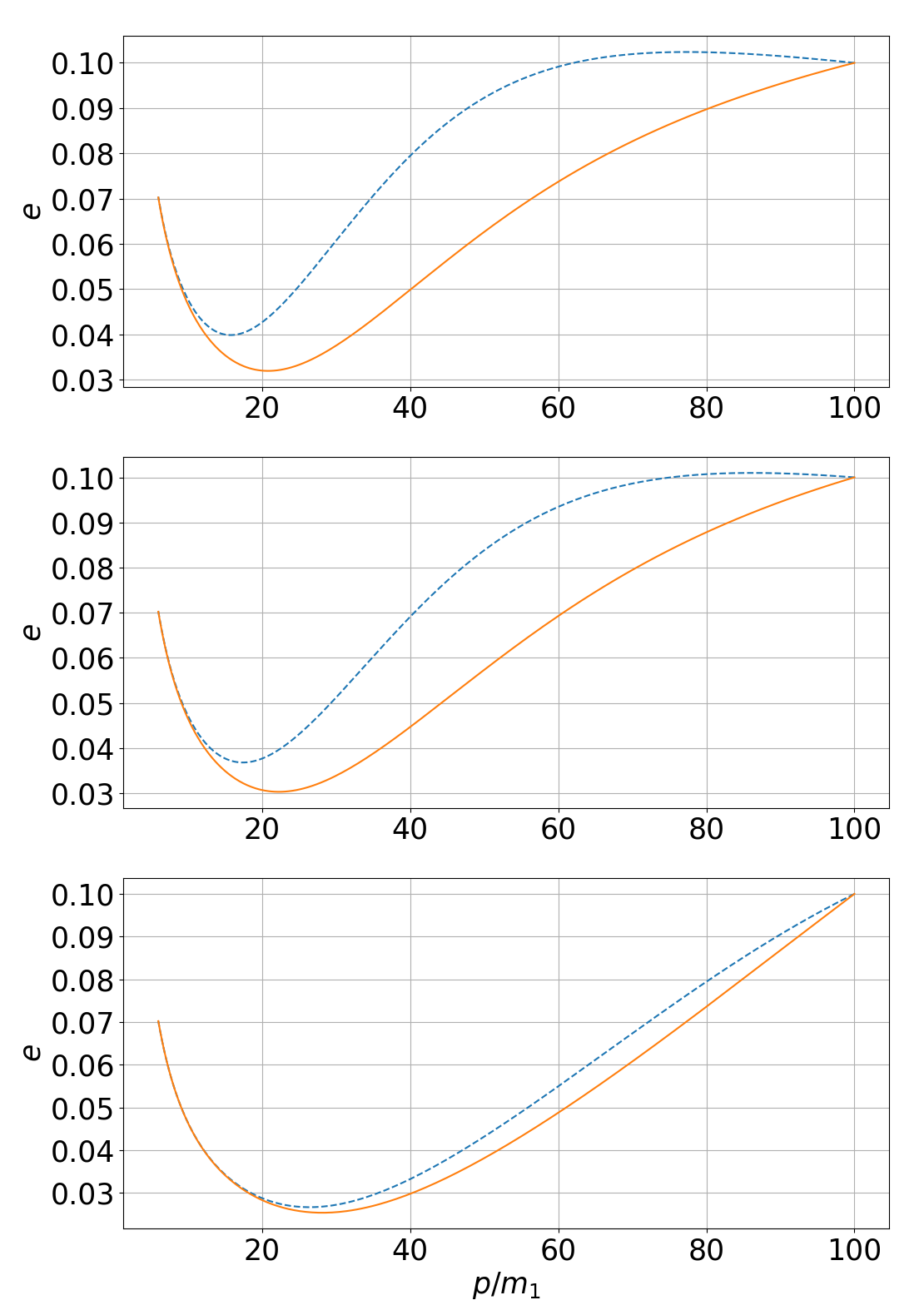}
\caption{The evolution of orbital eccentricity $e$ for a black hole spiraling towards a Kerr IMBH for a static halo (dashed) and a variable velocity halo (solid). The IMRI and the secondary mass values are $m_1 = 10^3M_{\odot}$, $m_1 = 1M_{\odot}$. The DM spike powers in the panels from top to bottom are $\alpha = [7/3, 9/4, 7/4]$.} 
\label{fig:fig7}
\end{figure}

\begin{figure}[t]
\centering
\includegraphics[width=0.99\linewidth]{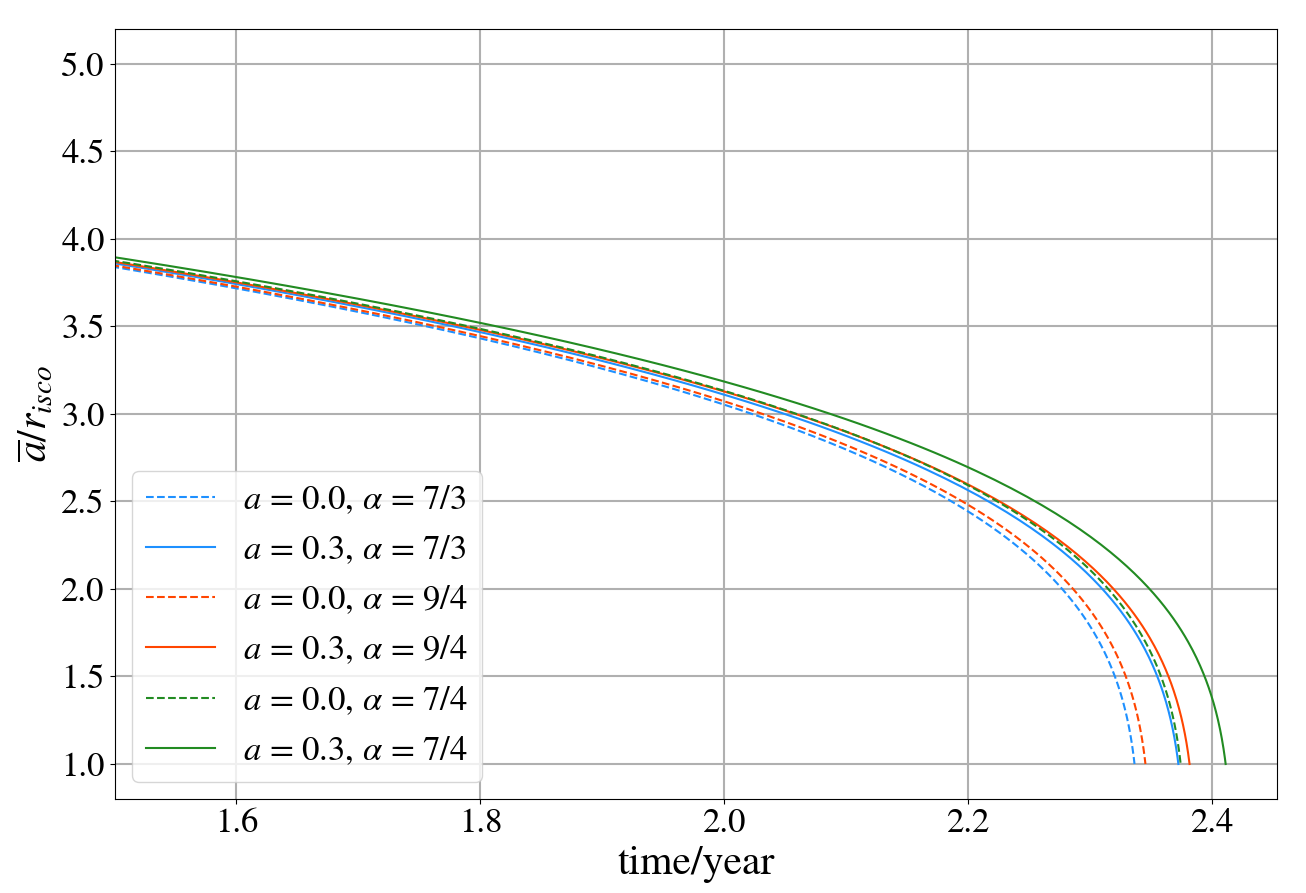}
\caption{
The evolution of the orbit's semi-major axis $\bar{a}$ for a non-rotating (dashed) and Kerr (solid) IMBHs as a function of time (in years). The initial eccentricity is $e_0=0.1$, the initial semi-major axis is $\bar{a}_0 = 5r_{\text{isco}}$,
and the DM spike density is $\rho_{\rm sp} = 5.448\times10^{17}$.
}
\label{fig:fig3}
\end{figure}

\begin{figure}[tt]
\centering
\includegraphics[width=1.08\linewidth]{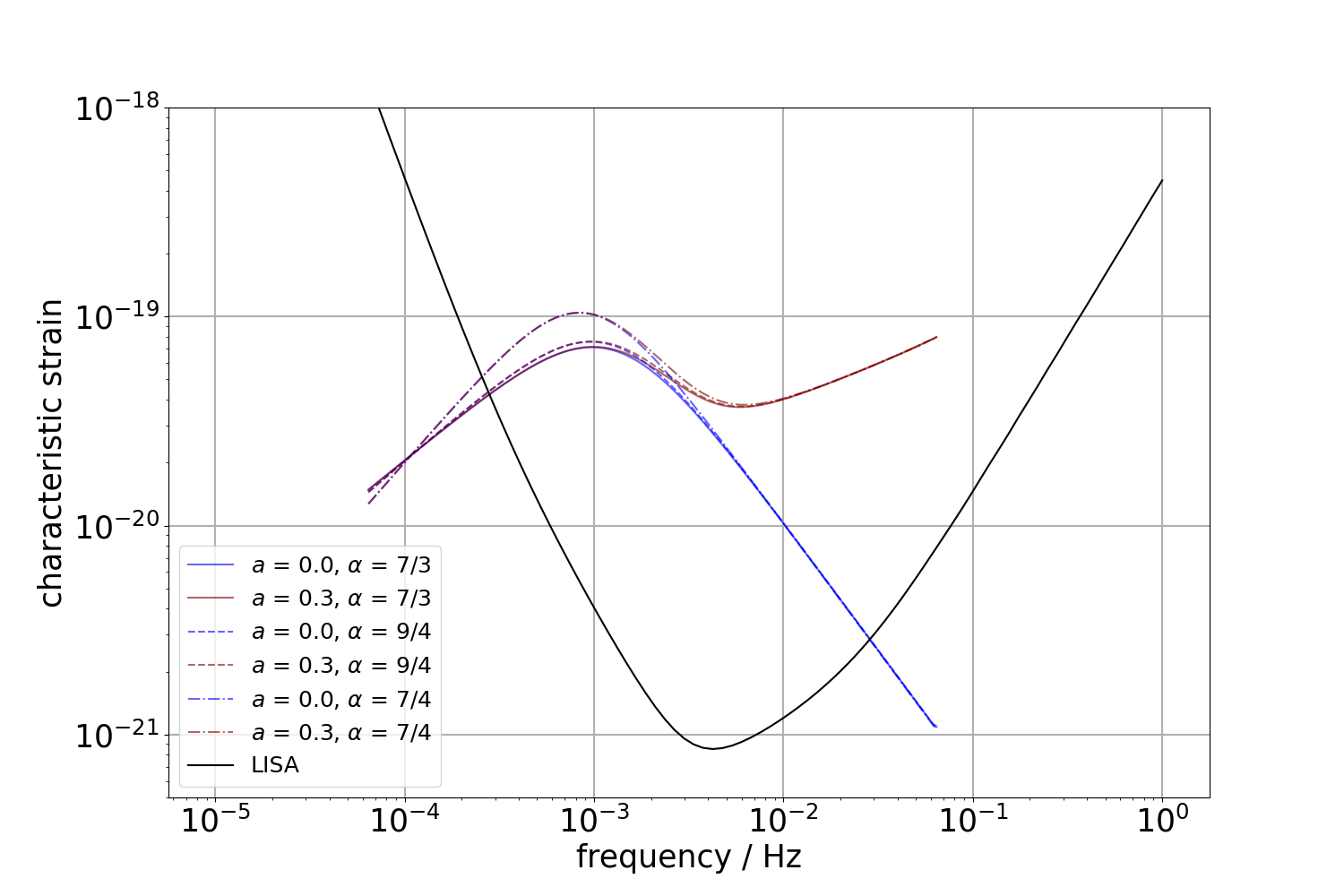}
\caption{The characteristic strain $h_{+}^{(3)}$ of the GW signal for the inspiral onto a non-rotating (blue lines) and Kerr (red lines) black hole, compared to LISA sensitivity curve (black line). The black hole masses are 
$m_1 = 10^5M_{\odot}$ and $m_2 = 10 M_{\odot}$, the luminosity distance is $D_L=100$ Mpc and the DM spike density is $\rho_{\rm sp} =  5.448\times10^{17}$.
Notice that the effect of angular momentum becomes pronounced at high frequencies while the effect of DM at low frequencies.}
\label{fig:fig4}
\end{figure}

We carried out numerical analysis of the evolution of the orbital parameters for the model outlined in the previous section
using a modified version of the publicly available code \texttt{IMRIpy} developed by N. Becker et al. \cite{Becker_2022, Becker:2022wlo,2023ascl.soft07018B}. Specifically, we used the ``kludge'' scheme in the modified code to generate the GW inspirals \cite{Babak:2006uv}.

We considered a system with $m_1 = 10^5M_{\odot}$, $m_2 = 10 M_{\odot}$ and $r_{\text{sp}} = 10^{-6} \ \text{pc}$. 
The initial semi-latus rectum has been taken as $p_0 = 100m_1$ and two different characteristic DM density parameters have been chosen, i.e. $\rho_{\text{sp}}= \left[5.448\times10^{15}, 5.448\times10^{17} \right]$ $M_{\odot}/$pc$^3$, for the two kinds of density profiles. For reference, we have also included the same simulations in vacuum.

Fig.~\ref{fig:fig1} shows the numerical results for the evolution of the eccentricity. The dashed lines represent inspirals onto a static Schwarzschild (i.e. $a = 0$) black hole, whereas the solid lines represent inspirals onto a Kerr black hole with the spin $a = 0.3$. 
As the semi-latus rectum decreases throughout the inspiral, the plot shows a temporal progression from right to left. Using three distinct choices of $\alpha =[7/3, \ 7/4, \ 9/4]$ in the density profile yields slightly varied outcomes, with the denser DM profiles producing results where the effects of DM are more pronounced. 
In the case of the NFW model ($\alpha =7/4$) the eccentricity decreases faster, while the SIDM ($\alpha =7/3$) and primordial black hole ($\alpha =9/4$) models have a slower descent.

As expected, during the inspiral onto a non-rotating black hole, 
in the weak-field region, the secondary object's orbit tends to circularize as a result of the influence of dynamical friction and gravitational radiation back-reaction. These results for non-rotating black holes align with the results previously obtained in \cite{Becker_2022}, indicating that when the stellar mass object approaches the IMBH situated in the center, its eccentricity diminishes, resulting in a more circular orbit. 
Our results show that for a DM profile with a relatively higher density of $\rho_{\text{sp}}=5.448\times10^{17}M_{\odot}/$pc$^3$, the eccentricity decrease is steeper. Notice that our simulation terminates when the value of the semi-major axis reaches $r_{\text{isco}} = 6m_1$ for a Schwarzschild BH.

On the other hand, the situation for inspirals onto a Kerr black hole is different. The deviation from circular orbit initially decrease  
in a manner analogous to that of a non-rotating black hole. Nevertheless, the eccentricity begins to rise when the semi-latus rectum reaches a value of approximately $p\lesssim30m_1$. This phenomenon arises due to the increased prominence of the spinning effects of the Kerr black hole on the spacetime when the secondary black hole approaches the central black hole. These findings are consistent with the results presented in \cite{Glampedakis_2002}.

The increase in eccentricity in the later stages of the inspiral is independent of the initial eccentricity, as can be seen from Fig.~\ref{fig:fig5}. Initial configurations with different eccentricities converge towards the same behavior once the semi-latus rectum becomes small enough $p\lesssim 15m_1$.

Subsequently, we examined the influence of the mass ratios and the DM particle velocities on the orbital parameter variations. The results for the evolution of the eccentricity for different mass ratios are shown in Fig. ~\ref{fig:fig6}. The decrease in eccentricity for smaller mass ratios is slower initially for $\alpha = 7/3$ and $9/4$. However, the behavior gets flipped for $\alpha = 7/4$. The distribution function is dependent on $m_1^{-\alpha}$, as demonstrated by the equation \eqref{eq:distr}. Consequently, with a simultaneous decrease in $m_1$ and an increase in $\alpha$, the distribution density increases, resulting in a relative increase in the dynamical friction force. As noted in \cite{Yue_2019,Becker_2022} the increase in dynamical friction force results in an eccentrification of the orbit due to the inverse dependence of the friction force on the square of $v_{\text{orb}}$. 

Fig. ~\ref{fig:fig7} illustrates the difference when considering the velocity distribution of the DM particles. It shows the eccentricity evolution for a single mass ratio $m_1/m_2 = 10^3 $ in a static halo (dashed) and a halo with variable velocity distribution of DM particles (solid). For $\alpha = 7/3$ and $9/4$, the eccentricity initially increases for the static halo in contrast to a variable velocity halo, matching the results in \cite{Yue_2019,Becker_2022}. All three cases show a further characteristic increase in eccentricity close to the ISCO.

The semi-major axis evolution for a Schwarzschild and a Kerr black hole in the presence of a DM halo is shown in Fig.~\ref{fig:fig3}. To establish a suitable scale, we choose the starting semi-major axis as $\bar{a}_0 = 5r_{\text{isco}}$. The plot demonstrates that the time required for inspiral onto a Kerr black hole is larger than that of the Schwarzschild case. 
Furthermore, the difference is mostly due to the central black hole angular momentum, with the impact due to the DM density profiles being qualitatively similar for all models. It can be noted that as the value of $\alpha$ increases, the inspiral time decreases. The resulting duration of the inspiral is then largely due to the growth in the orbital eccentricity when $a\neq 0$.

Finally, we examined whether the GW signals for the inspirals in the above models could be detected and distinguished by LISA. 
To analyze the possibility of detecting gravitational waves with the LISA detector, one can consider the dimensionless characteristic strain of the GW signal. The characteristic strain includes the effect of integrating an inspiralling signal. The signal-to-noise ratio is then defined as the ratio of characteristic strain and noise amplitude (which depends on the power spectral density function of the noise of the detector) integrated over the log of frequency (see \cite{Moore_2014} for a detailed discussion on GW sensitivity curves). Plotting the characteristic strain and the noise amplitude then allows one to easily check the detectability of a signal. The LISA power spectral density function is adopted from \cite{Robson_2019}. In Fig.~\ref{fig:fig4} we show the dimensionless characteristic strain of the wave in comparison with the sensitivity of LISA. The figure shows the characteristic strain for the 3rd harmonic of the Fourier-Bessel expansion. The waveforms for the 2nd harmonic are not shown, since they are not representative.

The plot clearly shows that the waveforms of inspiral onto a Kerr black hole are substantially different from the Schwarzschild case. The characteristic strain of a non-rotating black hole decreases monotonically with increasing frequency (ranging from $\sim 10^{-2}$ to $\sim 10^{-1}$ Hz), while in the Kerr case the characteristic strain exhibits the opposite trend, with the strain increasing with larger frequencies. 
The luminosity distance needs to be around $D_L = 100$ Mpc to be able to detect the presence of the DM halo and the signal drops below the noise for $D_L\gtrsim 500$ Mpc.

Significant differences appear in the waveforms for various $\alpha$ values as the DM density $\rho_{\rm sp}$ grows sufficiently large. 
The middle and bottom panels of Fig.~\ref{fig:fig4} show that the density of the DM environment plays a significant role in determining the frequency at which the characteristic strain peaks. These differences highlight the impact of the DM spike's density on the GW signal, providing a potential observational signature for the presence of DM halos around black holes.

\section{Conclusion}\label{conc}

We considered an IMRI system featuring a slowly rotating Kerr black hole, enveloped by a dark matter halo, at the core and a stellar-mass black hole orbiting around it. The inspiral of the stellar-mass object onto the central black hole generates gravitational waves, and measurement of the characteristic strain of the signal as a function of the frequency can provide information about the central black hole's angular momentum and the existence and properties of the dark matter halo surrounding it. To model the evolution of the orbital parameters and the characteristic strain of the system, we use the ``kludge'' method, which uses a relativistic treatment of the orbits while the fluxes are approximated at post-Newtonian orders. We emphasize that even though the kludge method does not produce entirely accurate results, it still captures the important features required for a qualitative understanding of EMRI/IMRI evolution \cite{Babak:2006uv}.

We have shown that the eccentricity initially decreases as the secondary object spirals towards the central black hole, and eventually increases again as the orbiting companion gets closer to the central black hole. We showed that the presence of dark matter of sufficiently large density affects the dynamics by causing an initial slower slope in the decrease of the eccentricity with respect to the vacuum case.

In fact, the increase in eccentricity at small distances due to non-vanishing angular momentum of the central black hole results in an increase in the characteristic strain signal of the emitted gravitational waves at larger frequencies, as opposed to the monotonic behavior produced by a non-rotating central black hole.

Also, the effects of dark matter on the orbital dynamics of the companion can be seen in a shift of the frequency corresponding to the maximum of the characteristic strain of the signal detected by gravitational wave experiments.
This maximum, appearing at lower frequencies, is related to the slower decrease in the orbital eccentricity with respect to the vacuum case and can provide information on the dark matter halo's density.

Future space-based detectors such as LISA will have the capability to detect gravitational radiation in the frequency range resulting from the capture of compact objects of stellar mass by massive black holes. If the luminosity distance is approximately $100$ Mpc, LISA (or alternatively, TianQin or Taiji) should be able to detect these waveforms and thus provide information on the properties of the central black hole and its surrounding environment \cite{Vazquez-Aceves:2022wmv}.

\section*{Acknowledgment}
The authors acknowledge funding support by Nazarbayev University under Faculty-development Competitive Research Grant program for 2022-2024, grant No. 11022021FD2926.

\bibliography{references}

\end{document}